\documentclass[runningheads]{llncs} 

\usepackage{booktabs} 
\usepackage{graphicx}

\usepackage{amsmath}
\usepackage{graphicx}
\usepackage{caption}
\usepackage{subfigure}
\usepackage{lipsum}
\usepackage{xcolor}

\usepackage[title]{appendix}

\usepackage{lipsum}

\newcommand\blfootnote[1]{%
  \begingroup
  \renewcommand\thefootnote{}\footnote{#1}%
  \addtocounter{footnote}{-1}%
  \endgroup
}

\newcommand{\rev}[1]{\textcolor{black}{#1}}

\begin{document}
\title{Share Withholding Attack in Blockchain Mining: Technical Report
\vspace{-0.15in}}

%
\author{Sang-Yoon Chang}%
\institute{University of Colorado Colorado Springs, Colorado Springs CO, USA\\
\email{schang2@uccs.edu}}



\maketitle \vspace{-0.1in}

\begin{abstract}
\vspace{-0.15in}
Cryptocurrency achieves distributed consensus using proof of work (PoW).  
Prior research in blockchain security identified financially incentivized attacks 
based on withholding blocks
which have the attacker compromise a victim pool and pose as a PoW contributor by submitting the shares (earning credit for mining) but withholding the blocks (no actual contributions to the pool).
We advance such threats to generate greater reward advantage to the attackers while undermining the other miners 
and introduce the share withholding attack (\rev{SWH}).
SWH withholds shares to increase the attacker's reward payout within the pool, in contrast to the prior threats withholding blocks, 
and rather builds on the block-withholding threats in order to exploit the information about the impending block submission timing, 
challenging the popularly established assumption that the block submission time is completely random and unknown to miners.
We 
analyze SWH's incentive compatibility and the vulnerability scope
by identifying the critical systems and environmental parameters which determine the attack's impact.
Our results show that SWH in conjunction with block withholding 
yield unfair reward advantage at the expense of the protocol-complying victim miners
and that a rational 
miner will selfishly launch SWH to maximize its reward profit.
We inform the blockchain and cryptocurrency research of the novel SWH threat and include the potential countermeasure directions to facilitate such research and development.
 \blfootnote{This technical report supplements a conference publication at SecureComm 2020~\cite{swh_2020}, which presents a shorter version of this work. To better highlight the differences, this technical report includes appendices which are excluded from the conference paper. More specifically, this technical report includes the following additional materials: more theoretical analyses and proofs, the more detailed Nash equilibrium analyses, the simulations with greater discussions about the setup and the SWH payout results, and the discussions about the potential countermeasures.} 
\end{abstract} 

\vspace{-0.25in}
\vspace{-0.25in}

\section{Introduction}
\label{sec:intro}
Blockchain
builds a distributed ledger 
and has emerged as the enabling technology for cryptocurrencies,
which generate and process the financial transactions without relying on a centralized authority such as a bank, 
e.g., Bitcoin~\cite{nakamoto2008} and 
Ethereum~\cite{buterin2014,wood2014}.
Cryptocurrencies operate in a permissionless environment lacking the pre-established trust in identities,
and the underlying distributed consensus protocols based on proof of work (PoW) enable
the nodes to agree on the ledger transactions 
by making the consensus fair with respect to the computational power (as opposed to the number of identities/votes);
the probability of finding a block and winning the corresponding reward is designed to be proportional to the computational power/hash rate by the PoW consensus protocol. 
Such PoW-based distributed consensus protocol is the most popular consensus protocol 
in the real-world blockchain implementations.
The miners participate in the PoW consensus to generate new currency and process the transactions
and are financially incentivized to do so by the block rewards, which are winnings from solving the probabilistic PoW computational puzzles.
To lower the variance of such reward income, the miners join and operate as mining pools
to share the computational power and the corresponding reward winnings.
Within a mining pool, to better estimate the individual miner members' contributions, mining pools use \emph{shares} which correspond to solving the same PoW computations as the blocks but with easier difficulty, providing greater number of samples for the contribution estimation.
If a block is found within the mining pool, instead of the miner finding the block getting the entire reward, the reward gets distributed across the shares so that the miners have lower variance in their reward earnings. 

Despite the consensus protocol indicating that the miner submits a block (a valid PoW solution) once found~\cite{nakamoto2008},
recent research in blockchain security
identified practical and relevant attacks
which have
the attacker withhold and control the timing of the block submission (including permanently withholding and discarding the block, as is in \rev{the classical block-withholding attack or BWH}) for unfair reward advantage over the protocol-complying strategy of immediately submitting the found block.
In such block-withholding threats,
the attacker compromises a victim mining pool and undermines the pool winnings by posing as a PoW contributor without honestly contributing;
while the reward winnings are shared in the \rev{victim} pool, the attacker additionally has a separate reward channel in its main pool/solo mining, in which it does not need to share the reward with others.
These attacks are in the forms of 
block-withholding attack (BWH), fork-after-withholding (FAW), and uncle-block attack (UBA).
FAW builds on selfish mining and BWH to advance and generalize BWH, 
and UBA further advances FAW by exploiting uncle blocks and making use of all the withheld blocks.
These threats 
are discussed in greater details in Section~\ref{sec:related}. 

In this paper, we advance the withholding-based attacks and introduce
the 
\emph{share-withholding} (SWH) attack.
SWH
withholds and delays the submission of the shares (as opposed to just the blocks)
to increase the reward payout within the victim mining pool.
In SWH, the misbehaving attacker exploits that it can gain some/probabilistic information about the impending block submission timing, 
\rev{which challenges the previous belief that block arrival timing is completely random to the miners thanks to the block arrival being a Poisson process~\cite{nakamoto2008}. 
Because knowing the block submission timing is critical for incentivizing SWH (as we show in Section~\ref{sec:swh_scope}), SWH builds on the aforementioned block-withholding threats, 
in which  
the attacker withholding and controlling the timing of the block submission opens opportunities for SWH gain. SWH further amplifies the reward gain beyond the state of the art block-withholding threats }
(where FAW and UBA already outperform protocol compliance and forgo the miner's dilemma)
by increasing the attacker reward at the expense of the other miners within the victim pool.
The additional reward gain is from the payout/contribution-estimation manipulation within the victim pool (which is different from the source of the reward gain for BWH/FAW/UBA as the block-withholding threats increase the reward of the main pool) as discussed in Section~\ref{subsec:difference_swh_bwh}.

Our analyses show significant gains and incentives for launching SWH even though
our reward/payout analyses is 
conservative in measuring the attacker's performances. 
For example, we use lower bounds and quantify the performances when the attacker loses the forking race and its fork-after-withheld block (which distinguishes FAW from BWH) does not become the main chain. 
SWH 
remains effective
even when the probability of the attacker's withheld block becoming main chain is zero (and the FAW attack reduces to the suboptimal BWH).
Furthermore, when uncle rewards are implemented (as in Ethereum~\cite{buterin2014,wood2014}), SWH further exploits the shares for even greater reward advantage than UBA or FAW.


Our work is generally applicable to all mining-based blockchains and to all rational miners.
We use formal modeling and analyses to identify the blockchain components which yield the SWH vulnerability, define the attack scope, and determine its impact and the impact dependency on the parameters.
Our model is driven by the real-world blockchain system designs and implementations and is applicable to all PoW-consensus-driven blockchains.
Throughout the paper, we construct the model and introduce additional complexities/parameters as they are used; the following sections often build on the previous sections and analyses.
Furthermore, in addition to sabotaging and undermining the other protocol-complying miners, our threat model supports a rational miner driven by its self profit
since we analyze the \rev{incentive compatibility. 
A mining strategy is \emph{incentive compatible} if its expected reward is greater than other known mining strategies, including protocol compliance, and
the miners driven by self profit would be incentivized to launch SWH} as long as they are uncooperative and willing to diverge from the given protocol.
Therefore, our threat model is stronger and more applicable than assuming only malicious and irrational miner which focuses only on disrupting the performances of the other victim miners.

\section{Background in Blockchain and Mining}
\label{sec:background}


The PoW consensus protocol participation is incentivized by financial rewards for generating a valid PoW which becomes the new block in the ledger.
The protocol participation is called \emph{mining} and the participants \emph{miners}
because the reward for finding the block include new currencies.
Only the miner which solves the PoW puzzle and finds the block the earliest\footnote{\rev{To determine which block was found the earliest can be a challenge in a distributed environment. To provide greater details about such resolution, we describe forking and how that can be resolved later in this section, and we describe uncle blocks/rewards adopted by Ethereum and newer cryptocurrencies which provide rewards to more than one miner in Section~\ref{sec:related}.}} wins the corresponding reward in that round (\rev{where each round increases the blockchain's block height by one}),
since the rest of the miners accept the newly found block and start a new round of mining by updating the chain with the found block.
The PoW consensus protocol is designed to be computationally fair, distributing the reward winning proportionally to the computational power of the miners in expectation. 
For example, assuming the protocol compliance of the miners, hundred miners, each of which has an equal hash rate of $x$ H/s, collectively earns the same reward amount as one miner with a hash rate of $100x$ H/s in expectation. 

Because a miner is competing with a global-scale group of other miners and the mining difficulty gets adjusted accordingly, solving a block is sporadic and of high variance.
%
To lower the variance and to get a more stable stream of reward income, miners form a pool to combine their computational power 
and share the corresponding mining rewards.
The increased computational power by pooling them together increases the occurrence of winning a block
and the corresponding reward gets split across the pool miners according to their computational contributions (which reward split within the pool is called \emph{payout}).
To estimate each miner's contributions, the mining pool samples more PoW solutions by using \emph{shares}, 
which correspond to solving the same computational puzzle with the same block header as the block but with easier difficulty.
The PoW solution corresponding to a share fixes \rev{fewer} number of bit 0's in the most significant bits of the hash output and therefore has a weaker constraint and a greater occurrence/probability than the PoW solution corresponding to a block.
In other words, if the block corresponds to finding a preimage/input $x$ which satisfies $H(x)<\tau_{\mbox{block}}$ where $\tau_{\mbox{block}}$ is the target threshold of the PoW puzzle, then the share corresponds to finding $x$ satisfying $H(x)<\tau_{\mbox{share}}$, where $\tau_{\mbox{share}}>\tau_{\mbox{block}}$,
and thus a block solution is/implies a share but a share is not necessarily a block.
To manage the mining pool, the \emph{pool manager} keeps track of the share count, registers/broadcasts the block upon its discovery, and distributes the reward-payout to the pool members according to their share submissions.
While optional, joining the mining pool to get a more stable, low-variance reward income is popular;
for example, in Bitcoin,
more than 89\% of the mining computation 
came from known mining pools~\cite{blockchain_info_hashrate}, which figure is a conservative lower bound because there are unidentified mining pools.

Due to the imperfect/asynchronous networking to submit and broadcast the blocks,
\emph{forking} occurs 
when two block solution propagations result in a collision (i.e., some nodes receive one block earlier than the other while the other nodes receive the other block first),
creating a disagreement/partition between the miners on which block was mined first and which to use for its impending round of mining.
Forking gets resolved by having the miners choose the longest chain (where the length of the chain can be measured by the number of blocks or the total difficulty),
e.g., if one partition finds a second block and propagates that to the other partition, then the miners in the other partition accept that chain which is one block longer than the one that they have been using.
\section{Related Work in Blockchain Mining Security}
\label{sec:related}



Following the consensus protocol of timely block submissions has been believed to be incentive compatible, as the block submission monotonically increases the reward at the time of the submission. 
However, given the same computational power, more sophisticated attacks emerged to further increase the mining reward 
by operating against the protocol and controlling/delaying the timing of block submission, 
including permanently withholding the submission in certain situations.
\emph{Selfish mining} withholds a block so that the miner can get a heads-start on computing the next block and have the rest of the miners discard and switch from the blocks that they were mining~\cite{eyal2013,gervais2016}. 
However, the confirmation mechanism, introduced by Bitcoin and inherited by most blockchain implementations, resists selfish mining by waiting until multiple blocks are further mined after a block is found,
decreasing the probability of successful selfish mining exponentially with the number of blocks needed for confirmation~\cite{nakamoto2008}. 

Against mining pools, there are even more advanced threats. 
The \emph{block-withholding (BWH) attack} 
withholds the mined block in the victim pools in order to increase the attacker's overall reward 
(and specifically that of the main pool) at the expense of the rest of the victim pool members~\cite{rosenfeld2011}.
In BWH, to sabotage the victim mining pool, the attacker simply never submits the found block while submitting the shares.
As a consequence, the attacker still reaps the benefits from submitting the shares to the victim pool (pretending to contribute and getting the credit for it) while never actually contributing to the pool (since it never submits the actual block solution which will benefit the victim pool).

\emph{Fork-after-withholding (FAW) attack}~\cite{kwon2017} 
builds on selfish mining and BWH but creates intentional forks in cases when there is a block being broadcasted by a third-party pool (with which the attacker has no association).
In other words, while always submitting the shares to gain greater payout on the victim pool, the attacker withholds the found block and either discards it (if the attacker's main pool finds another block or if another miner from the victim pool finds a block) or submits it only when there is another competing block that is being propagated by another third-party pool (creating an intentional fork).
Unlike BWH yielding no reward if the third-party pool submits a block, FAW causes a fork to yield a statistically positive reward (i.e., the attacker wins the forking race sometimes) when the third-party finds and submits a block. 
FAW is significant because it forgoes the miner's dilemma 
(motivating the pools to cooperate with each other)~\cite{eyal2015},
and there is a real incentive (unfair reward gain) to launch FAW for rational miners.

\emph{Uncle-block attack (UBA)}~\cite{UBA2019} exploits the uncle blocks \rev{which provide partial rewards for the blocks which got mined but did not become the main chain, e.g., Ethereum to provide greater fairness in the miners' networking conditions. UBA} builds on FAW, inheriting the attacker-desirable properties, but advances it to generate greater unfair rewards to the attacker by making use of all the withheld blocks (in contrast to FAW discarding some withheld blocks) and by making it relevant and impactful even when the attacker has suboptimal networking and loses the forking race (in which case, FAW reduces to the suboptimal BWH).

The research and development for defending against such block-withholding threats and aligning the incentives to protocol compliance is ongoing, e.g.,~\cite{buterin2019,bag2016,silenttimestamping2019,rosenfeld2011}.
Rosenfeld~\cite{rosenfeld2011} (which introduced the aforementioned BWH attack) 
also introduces payout algorithms to build resistance against pool-hopping attacks,
in which the attackers dynamically hop between pools for increased reward. 
His seminal work and the analyses of the payout schemes are widely adopted in the modern-day mining pool practices. 
Unfortunately, we later see in Section~\ref{sec:swh_scope} that one of Rosenfeld's main inventions for defeating pool hopping (decaying payout, as we call it) yields a critical vulnerability for our novel threat of SWH.  

\section{Threat Model}
\label{sec:threat_model}

We define \emph{honest miners} to be cooperative and follow the consensus protocol \rev{and the mining pool protocol, including the timely block and share submissions}. 
The other non-honest miners can launch other mining strategies (e.g., those described in Section~\ref{sec:related} or SWH) and are rational (choose the strategy achieving greater reward). 
More specifically, the non-honest miners intelligently control the timing of the PoW solution submissions (blocks or shares). 
If the non-honest and rational miners violate the consensus protocol, 
we call them \emph{attackers}
as they operate at the expense of the other honest miners. 

\begin{figure}[t]
\begin{center}
\includegraphics[width=0.40\columnwidth]{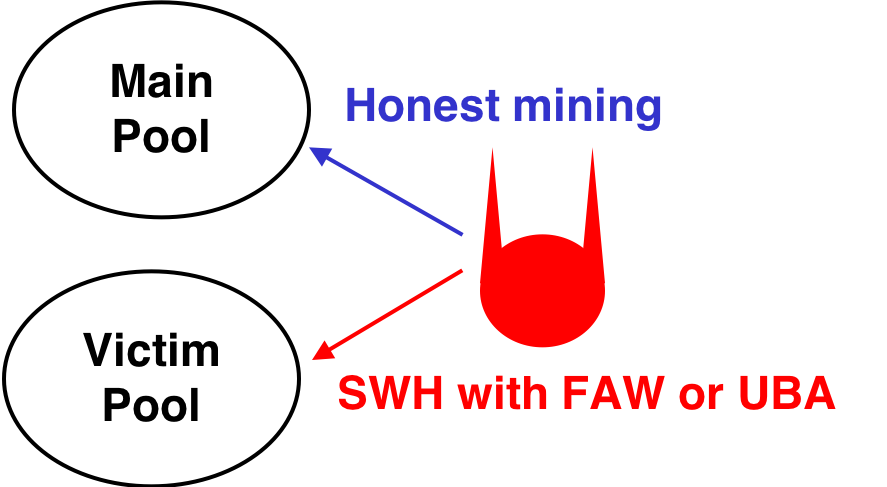}
\caption{The SWH setup is the same as BWH, FAW, or UBA and includes compromising the victim pool. The attacker splits its computational power between mining honestly in the main pool and infiltrating the victim pool to launch the mining threats. 
However, while FAW or UBA increase the reward for the main pool from such infiltration at the expense of the victim pool as a whole, SWH increases the attacker's reward split within the victim pool. \vspace{-0.3in}}
\label{fig:main_pool_victim_pool}
\end{center}
\end{figure}

\subsection{The Same Setup for SWH as BWH} 
\label{subsec:same_setup}

We investigate the attacks launched by the miners.
%
We assume the threat model of BWH (the same as FAW and UBA), in which the attacker compromises multiple pools and separates the main pool vs. the victim pool\rev{, as depicted in Figure~\ref{fig:main_pool_victim_pool}}.
The attacker behaves honestly in the \emph{main pool} while it can launch an attack by diverging from the protocol in the \emph{victim pool}\rev{, i.e., the attacker compromises a victim pool by joining the pool with the intention of launching an attack}.
As a realistic setup and for simplicity, 
we assume that the main pool is comprised of the attacker only (equivalent to solo mining);
the attacker shares the reward winnings in the victim pool while there is no sharing and the attacker takes all the rewards in the main pool.
Our model also generalizes to the case of multiple pools/miners (e.g., the main pool or the victim pool can be a collection of mining pools) \rev{as long as those comprising the main pool are under the attacker control
since the PoW consensus is designed to be power-fair} as opposed to identity-fair, \rev{as is described in Section~\ref{sec:background}} and captured in our mining-game model in Section~\ref{sec:mining_game}.

The attacker setup is realistic 
since cryptocurrencies operate in \rev{permissionless} environments with loose control of identities and is designed for anonymity. 
For example, in 2014, Eligius became a victim pool of BWH attack and lost 300 BTC\rev{~\cite{eligius_bwh}},
which got detected because the attacker only used two accounts for the attack (which resulted in the detection of abnormal behavior where the accounts submitted only shares but no blocks for too long of a time).
\rev{Combining such attack with Sybil attack (or just simply following Nakamoto's suggestion to use new accounts for new transactions for anonymity~\cite{nakamoto2008}) instead of just using two accounts} would have made such detection significantly more difficult. 


\section{Share Withholding Attack}
\label{subsec:difference_swh_bwh}

Share withholding attack (SWH) withholds shares in order to 
manipulate the intra-pool payout (shifting the share-based payout and the reward distribution within the pool), 
which is in contrast to the prior miner threats withholding blocks in Section~\ref{sec:related}
increasing the probability of winning in the attacker-main pool (at the expense of that of the compromised victim pool). 
%
%
SWH is therefore fundamentally different from the threats withholding blocks and can be launched separately with 
UBA/FAW/BWH.
However, SWH provides reward gain (incentive compatible and relevant to rational miners) if the attacker has some knowledge of the block submission timing as we will see in Section~\ref{sec:swh_scope}.
While there can be other cases to fulfill such condition requirement, we identify the block-withholding-based threats as practical cases where the attacker fulfills the requirement of having knowledge of the block submission time (since the attacker has control over the timing of the withheld blocks). 
\rev{More specifically, when the attacker launches SWH with FAW or UBA, it submits its shares right before its withheld blocks.} 
Therefore, we introduce SWH not only by itself but also in conjunction with the block-withholding-based attacks of UBA/FAW/BWH.  
When analyzing the impact of SWH in Section~\ref{sec:swh_reward} and Section~\ref{sec:sim}, we focus more on UBA and FAW because UBA generalizes FAW 
and FAW generalizes BWH; UBA is more advanced than FAW and FAW more advanced than BWH where, in both cases, the former strategy can be reduced to the latter in certain mining environments
(e.g., when uncle reward is zero, UBA gets reduced to FAW). 

\section{Mining Game}
\label{sec:mining_game}

This section builds on and adapts the models in 
the prior literature analyzing the withholding-based attacks, e.g.,~\cite{kwon2017,luu2015,UBA2019}. 
On the other hand, 
Section~\ref{sec:pool_game_analyses} and Section~\ref{sec:swh_reward}
extend the model to introduce a more complex model and a framework to analyze the intra-pool dynamics and SWH. 
As discussed in Section~\ref{subsec:difference_swh_bwh}, 
while block withholding increases the main pool reward and focuses on the inter-pool dynamics (this section), 
SWH 
manipulates the share-driven reward payouts within the pool 
(Section~\ref{sec:pool_game_analyses} and Section~\ref{sec:swh_reward}).

\subsection{Mining and Computational Power Model}
\label{subsec:computational_power_model}

To investigate the incentive compatibility of the attacks, we model the mining game between the miners and quantify the expected reward.
The expected reward depends on the miner's computational power, and we normalize the following variables with respect to the total miner network's computational power,
e.g., the entire miner network has a computational power of 1.
The attacker's computational power is $\alpha$ (where $0 \leq \alpha \leq 1-\beta$) while the victim pool's mining power excluding the attacker's power is $\beta$ (where $0 \leq \beta \leq 1-\alpha$). 
The power of the other pools/miners outside of the attacker and the victim pool is therefore $1-\alpha-\beta$. 
\rev{Building on the attacker setup in Section~\ref{subsec:same_setup} and in Figure~\ref{fig:main_pool_victim_pool},} the attacker splits its power between its main pool (honest mining) and the victim pool 
(possibly adopting mining attack strategies to increase the attacker's reward at the expense of the fellow miners in the victim pool),
and the fraction of the attacker's power for infiltration of the victim pool is $\tau$ (where $0 \leq \tau \leq 1$).
Therefore, the attacker's power on the victim pool is $\tau \cdot \alpha$, and the total mining power on the victim pool is $\tau \alpha + \beta$ even though the attacker's power does not fully contribute to the pool earning reward. 
For example, in the simpler block-withholding attack of BWH, the attacker does not submit block at all in the victim pool so the actual power contributing to block earnings of the pool is only $\beta$, while the attacker still earns the reward/credit through share submissions and the reward earning gets split by ~$\tau \alpha + \beta$. 


A miner's expected reward is denoted with $R$.
For example, if an attacker chooses to behave honestly (one of its possible choices), its expected reward ($R_{\mbox{honest}}$) is proportional to its computational power by the design of the PoW consensus and the mining pools,
\begin{eqnarray}
R_{\mbox{honest}} = \alpha
\label{eqn:reward_honest}
\end{eqnarray}

The following summarizes the variables for analyzing block withholding threats. 
\newline \indent \indent $\alpha$: \indent Attacker's computational power 
 \newline \indent \indent $\beta$: \indent Computational power of the victim pool 
\newline \indent \indent $\tau$: \indent Fraction of attacker's power for infiltration of victim 
\newline \indent \indent $c$: \indent Probability that the attacker wins the reward given that there is a fork (collision with another block propagation) 

\subsection{BWH, FAW, and UBA Analyses} 
\label{subsec:faw_recap}

To provide baselines and examples of the use of our model in Section~\ref{subsec:computational_power_model}, we analyze the expected reward of BWH and FAW.
This section adapts the prior work in the block-withholding-based threats~\cite{kwon2017,luu2015,UBA2019}, 
and we only extract the parts most relevant to our work in this section. 

For BWH, the attacker has two possible events for earning a positive reward (in other events, the attacker earns zero reward). 
The first event is when the attacker finds a block in its honest-mining main pool 
($A$) 
while the second event corresponds to when another miner from the victim pool, not the attacker, finds a block ($B$).
Because the probability of winning a block is proportional to the computational power spent on mining the block and because $1- \tau \alpha$ amount of power from all miners actually contributes to finding the block and ending the round (the attacker uses $\tau \alpha$ to only submit shares while withholding the blocks), the probability of $A$ is $\frac{(1-\tau) \alpha}{1-\tau\alpha}$ and 
the probability of $B$ is $\frac{\beta}{1-\tau \alpha}$.
Assuming negligible probability for natural forking, 
the expected reward for block-withholding attack ($R_{\mbox{BWH}}$) is: 
\begin{eqnarray}
R_{\mbox{BWH}} & = & \mbox{E}[R | A] \cdot \mbox{Pr}(A) + \mbox{E}[R | B] \cdot \mbox{Pr}(B) \nonumber \\
& = & \frac{(1-\tau) \alpha}{1-\tau\alpha} + \frac{\tau \alpha}{\beta + \tau \alpha} \cdot \frac{\beta}{1-\tau \alpha}
\label{eqn:reward_bwh}
\end{eqnarray}

The FAW attack builds on the block-withholding attack but provides an extra channel for attacker reward. 
In addition to the events $A$ and $B$, the attacker can earn reward by broadcasting the withheld block when a third-party miner outside of the attacker-involved main pool and victim pool finds a block, causing a fork and hence the name fork-after-withholding (FAW). 
This event of the attacker finding a block \emph{and} a third-party miner finding a block is $C$. 
The expected reward for FAW attack ($R_{\mbox{FAW}}$) is: 
{\footnotesize 
\begin{eqnarray}
R_{\mbox{FAW}} 
& = & \frac{(1-\tau) \alpha}{1-\tau\alpha} 
+ \frac{\tau \alpha}{\beta + \tau \alpha} \left( \frac{\beta}{1-\tau \alpha} + c \tau \alpha \frac{1-\alpha-\beta}{1-\tau \alpha} \right)
\label{eqn:reward_faw}
\end{eqnarray}
}

\rev{Appendix~\ref{appendix:uba_reward} also includes the reward analysis for UBA.}
We summarize the three events which yield the attacker positive rewards, as we also use them for our analyses of the rewards for the other attacks: 
\newline \indent \indent $A$: \indent Attacker's main pool finds a block
 \newline \indent \indent $B$: \indent Another miner from the victim pool finds a block
\newline \indent \indent $C$: \indent Third-party miner, not affiliated to attacker, finds a block 

\section{Mining Pool Game and SWH Scope Analyses}
\label{sec:pool_game_analyses}


While the mining game model in Section~\ref{subsec:computational_power_model} characterized the miners' activities at the inter-pool level, 
SWH 
requires greater details in the modeling of the intra-pool operations at the share/block submission level;
the model needs to capture the individual submissions of the blocks/shares and the reward split distribution within the mining pool (which we call \emph{payout}).  
To support such model, in this section, we introduce the mining pool game and model (Section~\ref{subsec:mining_pool_model})
and use it to analyze the SWH feasibility and scope (Section~\ref{sec:swh_scope}). 
Afterward, we analyze the SWH performance (Section~\ref{sec:swh_reward}). 
Section~\ref{sec:swh_reward} builds on Section~\ref{sec:pool_game_analyses}, 
focusing on the SWH-vulnerable scope, and provides greater details and increasing complexity in the model and analyses.
The variables are introduced as they are used, for example, Section~\ref{sec:swh_reward} focuses on the decaying-payout scheme (which is one of the two types of payout-schemes in Section~\ref{sec:swh_scope}). 


\subsection{Mining Pool and Share Model: $f$ and $\vec{s}$}
\label{subsec:mining_pool_model}

%
%
In a mining pool, there are $n$ miners, each of which is denoted with $m \in \{1,2,...,n\}$.
For example, if $n=1$, then $m \in \{1\}$ and it is solo mining (e.g., the attacker's main pool).
We characterize the $i$-th share submission ($\vec{s_i}$) using a pair of random variables:
the index of the member who submitted the $i$-th share $m_i$ 
and the time of submission $t_i$ (recorded by the mining pool manager). 
In other words, $\vec{s_i} = (m_i, t_i)$.  
The collection of these share submissions over time is denoted with $\vec{s}$. In other words, $\vec{s} = \{\vec{s_i}\}_i$ is the share history and $\vec{s}$ grows with the number of share submissions.
For example, from the beginning of the round, if the member 4 submitted the first share at time 6 and the member 10 shared the second share at time 9.5, then $\vec{s} = ((4,6),(10,9.5))$ if the second share is the most recent share.
The share list $\vec{s}$ is \rev{an implicitly ordered list in the share's time order (i.e., the shares get added at the end as they are found)} and continues to grow until the block is found and submitted (the end of the round), after which $\vec{s}$ resets to $\emptyset$.
When a new $i+1$-th share arrives, $\vec{s}$ with $i$ submissions/elements gets updated by $\vec{s} || (m_{i+1},t_{i+1})$ where $\vec{a}||\vec{b}$ denotes the concatenation of $\vec{a}$ and $\vec{b}$.
We assume that share rules and validity are enforced 
so that an invalid share submission (which gets rejected by the mining pool system) is equivalent to no submission.

The \emph{payout} to determine how the pool reward gets split between the pool members (as opposed to the final reward which accounts for the rewards from the other pools as well) depends on the share submissions.
To model the payout and generalize it to different payout schemes (how to divide the reward between the share submissions), we introduce
the payout function $f$ which divides the reward using the share submissions 
and produces a vector
in which the $i$-th element corresponds to 
the payout to the corresponding submitter $m_i$.
In other words, $f: \vec{s} \rightarrow \vec{R}$, i.e., $f$ uses the share history ($\vec{s}$) to generate the reward for each miners ($\vec{R}$)
where $\vec{R}$ is either a zero vector if the pool does not win the reward
or a vector with non-negative scalar elements with a size of $n$ and $\| \vec{R} \|_1 = 1$ (the element adds up to one so that each element corresponds to the fraction of the reward winnings).
From $f$, $f_j$ takes the j-th element of $\vec{R}$ and derives the reward for miner $j$, i.e., $f_j: \vec{s} \rightarrow R_j$.
For example, for Pay-Per-Share (PPS) scheme, $f_j$ counts for the number of shares submitted for each miner member $j$ 
and, for Proportional payout scheme, $f_j$ counts for the number of shares for each miner $j$ and divides it by the total number of shares. 

%
\subsection{SWH Vulnerability Scope}
\label{sec:swh_scope}

SWH is based on withholding shares and delaying their submissions (in contrast to block-withholding-based attacks which control block submissions). 
While we study the attack's implementation and impact in Section~\ref{sec:swh_reward}, 
we first investigate if delaying shares can be relevant and incentive-compatible 
and establish the vulnerability scope in which the share-withholding attack is relevant 
in this section. 
We analyze how the payout function $f$ (distributing the pool reward to the members) plays a critical role in determining the system's vulnerability against SWH. 

\subsubsection{Payout Scheme Definition}

The payout function $f$ is designed to increase the reward for a miner when it submits a share.
We define $f$ to be \emph{unilaterally increasing} if, given any share history $\vec{s}$, the submission of a share \rev{monotonically increases (i.e., either increases or remains constant as its input increases)} the payout of the submitter of that share. 
\rev{For any miner $j$, $f$ unilaterally increases with $j$'s share submission if any new share submission by $j$ at time $t$, corresponding to $(j,t)$ in the share list, monotonically increases $j$'s payout.}

\begin{definition}
$f$ unilaterally increases the payout with shares if $\; f_j (\vec{s}||(j,t)) - f_j(\vec{s}) \geq 0, \; \forall \vec{s}, \; \forall t, \; \forall j$.
\end{definition}

The real-world mining pools implement $f$ so that it unilaterally increases with a share,
and any share submission does not harm the subject miner's reward earnings. 


$f$ implementation is divided into two classes, which are mutually exclusive and cover all $f$ implementations. 
The first class of $f$ called fixed payout corresponds to when the payout for each share submission remains the same in time (as long as the share list remains the same)
and the other class called decaying payout corresponds to when the share payout gets decayed over time \rev{from the time of its submission}.
We say that the share has a \emph{fixed payout} if the time of the share submission does not affect the share attribution to the payout as long as the share list $\vec{s}$ remains constant (e.g., 
a new submission changes the share list $\vec{s}$ and can further spread the reward between the submissions).
Implied in the fixed-payout definition is that the share remains valid and stays within the same round because $\vec{s}$ remains the same; 
this is an important clarification for the SWH attack because there is a risk in delaying the share submission and losing its payout value (which occurs if any other miner found a block and the pool moves on to a new block header/round making the attacker's share stale). 

\begin{definition}
$f$ is a \emph{fixed payout} scheme if $ \; f_j (\vec{s}||(j,t+\Delta t)) - f_j (\vec{s}||(j,t)) = 0, \; \forall t, \; \forall \Delta t, \; \forall j, \forall \vec{s}$. 
\end{definition}

In real-world $f$ implementations, Proportional, Pay-Per-Share (PPS), Pay-Per-Last-N-Shares (PPLNS) fall within the fixed-payout schemes\rev{~\cite{rosenfeld2011}}.
In contrast to PPS, the share payout can vary as there are new submissions,
e.g., by having the payout 
depend on the number of shares and the size of $\vec{s}$ size (which may grow in time), as in the Proportional payout scheme.
Even in such cases, the share's payout attribution remains the same as long as the share is submitted within the same round.

On the other hand, other schemes for $f$ has a \emph{decaying payout}\rev{~\cite{rosenfeld2011}}, i.e.,
a share's payout value decays in its value in time,
and the more recent share submissions have a larger payout than an older share submission.

\begin{definition}
$f$ is a \emph{decaying payout} scheme if $ \; f_j (\vec{s}||(j,t+\Delta t)) - f_j (\vec{s}||(j,t)) > 0, \; \forall t, \;  \forall \Delta t, \; \forall j, \forall \vec{s}$.
\end{definition}

In real-world $f$ implementations, score-based/Slush's payout and Geometric payout fall under decaying-payout schemes.


%

\subsubsection{Fixed Payout Case}
We show that the attacker has no incentive to conduct share-withholding attack if $f$ has a fixed payout.
In other words, the attacker's choosing $\Delta t>0$ does not increase its payout.

\begin{theorem}
Attacker has no reward gain 
to delay its share if the payout scheme $f$ unilaterally increases the attacker's payout and has a fixed payout.
\end{theorem}
\begin{proof}
Suppose the attacker $j$, for any $j$, finds a share at time $t$ and it is valid at the time.
It is sufficient to show that, given any $\vec{s}$ and any $\Delta t >0$, the expected payout is less than that corresponding to $\Delta t =0$.
By definition of $f$ being fixed payout,
$j$'s payout for submitting the share at time $t+\Delta t$, $ f_j (\vec{s}||(j,t+\Delta t))$
is either $ f_j (\vec{s}||(j,t+\Delta t)) = f_j (\vec{s}||(j,t))$ or $ f_j (\vec{s}||(j,t+\Delta t)) = f_j(\vec{s}) $,
the latter of which is smaller than $f_j (\vec{s}||(j,t))$ because $f$ unilaterally increases with $j$'s share.
Since $\mbox{E}[f_j(\vec{s}||(j,t))]$ is a linear combination of the payouts corresponding to the two events (which partitions and comprises the entire possibilities - the share either remains valid or becomes invalid, e.g., no longer valid block header because it is a new round) by law of total probability,
$\mbox{E}[f_j(\vec{s}||(j,t))] \geq \mbox{E}[f_j(\vec{s}||(j,t+\Delta t))]$.
\end{proof}

\subsubsection{Decaying Payout Case}
Suppose the mining pool system uses $f$ such that it unilaterally increases the payout for every miners and has a decaying payout.
Given such $f$, the attacker has an incentive to conduct SWH if
the attacker knows the block submission time (the end of the round) ahead of time.
The knowledge of the block-submission timing is critical for SWH, as we will see in this section. 
While there can be other cases where such assumption holds and our analyses still applicable, 
we identify a concrete case when the attacker does have the block-submission time information in FAW/UBA. 
In FAW or UBA, an attacker knows the block submission timing and can launch SWH, 
because an attacker having found and withholding the block has the knowledge of the block submission time. 
The attacker can further increase its knowledge of the block submission time in advance by combining the withholding with networking-based attacks~\cite{apostolaki2017,ekparinya2018,heilman2015}.
Therefore, we focus on 
an attacker launching SWH in conjunction with FAW/UBA and analyze the SWH attack impact in such case in Section~\ref{subsec:swh_faw_reward}.

We first analyze the case when the attacker has the \emph{full information} about when the block will get submitted
(i.e., the attacker has the correct information about the block submission timing all the time).
Then, we build on the full-information case to analyze the case when the attacker has the information sometimes and those cases occur with a non-zero probability 
(we later introduce $c'$ for such probability when incorporating share-withholding attack with FAW attack in Section~\ref{sec:swh_reward}).

\begin{lemma}
Given a payout scheme $f$ which unilaterally increases the attacker's payout and has a decaying payout, attacker has a positive reward gain 
to delay its share if the attacker has full information about when the block will get submitted. 
\label{lemma:decaying_reward_complete_info}
\end{lemma}
\begin{proof}
Suppose the share is found at time $t$ and the block is found/submitted at $t_B$, known to the attacker.
Let's prove by contradiction.
Assume that the attacker choosing $\Delta t = 0$ maximizes the expected payout, which is $\mbox{E}[f_j(\vec{s}||(j,t+\Delta t))] = \mbox{E}[f_j(\vec{s}||(j,t))]$.
Let attacker withhold the share and choose $\Delta t = t_B - t - \epsilon, \; \forall \epsilon >0$.
Then, by definition of $f$ having decaying payout,
$ \; f_j (\vec{s}||(j,t+t_B - t - \epsilon)) = f_j (\vec{s}||(j,t_B - \epsilon)) > f_j (\vec{s}||(j,t))$.
Therefore, the attacker is incentivized to withhold shares.
\end{proof}

\begin{theorem}
Given a payout scheme $f$ which unilaterally increases the attacker's payout and has a decaying payout, attacker has a positive reward gain 
to delay its share if the attacker has the information about when the block will get submitted with a non-zero probability.
\end{theorem}
\begin{proof}
Lemma~\ref{lemma:decaying_reward_complete_info} states that the attacker is incentivized to launch share-withholding attack if the attacker has the full information of the block submission timing.
The attacker can either withhold a share or promptly submit it.
If the attacker only withholds shares when it has the knowledge of the block submission timing,
then the expected payout of such strategy is a linear combination between the payout when share-withholding attack with perfect knowledge
and the payout with no share-withholding,
which is greater than the latter (no share-withholding) if there are cases when the attacker has the knowledge with non-zero probability.
\end{proof}


\subsection{SWH Vulnerable Scope in Real-World Practice}
\label{subsec:real_world_implementations}

The payout function $f$ is critical in determining whether the mining pool is vulnerable against SWH. 
More specifically, if $f$ is of decaying payout (as opposed to fixed payout), then the mining pool is vulnerable against SWH. 
In our investigation, more than 12\% of the miner computations in Bitcoin (measured by the computational power) use decaying-payout scheme and are vulnerable against SWH; 
this is a conservative figure since it only counts the mining pools whose designs are known, 
and there are many pools which obscure their payout functions. 
For example, SlushPool (well regarded in cryptocurrency community thanks to its transparency and state of the art security practices) is within the vulnerable against SWH.

\section{SWH Payout and Reward} 
\label{sec:swh_reward}

In SWH, the attacker withholds shares for submission until the blocks get submitted in order to increase the payout of the shares. 
Since the attacker blindly delaying shares risks the shares becoming stale/outdated lowering its payout, we analyze two cases: the case when the attacker has the full information about the block submission timing
and the case when the attacker combines share-withholding with block-withholding-based attacks (where the attacker has probabilistic control over the block submission timings).
We build on the analyses of the former case for the analyses of the latter case
to provide a concrete scenario where the attacker has the information of the block submission timing because it actively controls the submission timing. 

SWH is incentivized because the withholding increases the attackers' expected payout.
While Section~\ref{sec:swh_scope} established the vulnerability scope of SWH
(SWH is incentivized when $f$ has a decaying payout and when the attacker knows the block submission timing), 
we analyze the attack impact in the attacker's payout/reward 
while assuming that the victim pool adopts decaying payout in this section. 
We first study SWH payout (within a pool)
and then incorporate that to the reward analyses of SWH coupled with FAW and UBA (not only incorporating the payout from the victim pool but also accounting for other reward channels from other pools, e.g., the attacker's main pool). 


\subsection{Decaying Payout Model}
\label{subsec:model_decaying_payout}

We build on the mining and computational power model in Section~\ref{subsec:computational_power_model} and the mining pool model in Section~\ref{subsec:mining_pool_model}.
This section focuses on the additions to the model to analyze the share-withholding attack.
More specifically, it introduces additional parameters to better describe the decaying-payout function $f$, as opposed to leaving it as an abstract function as in Section~\ref{subsec:mining_pool_model}.

As described in Section~\ref{sec:swh_scope}, a payout scheme with a fixed payout is not vulnerable to SWH by a rational, incentive-driven attacker.
However, a decaying-payout scheme does incentivize a rational attacker to withhold shares and delay their submissions.
A popular implementation of a payout with a decaying payout is in the form of \emph{scores}, which quantify the share values for the payout and have each share experience exponential decay in its payout value. 
Such scoring-based payout is used in Slush, geometric, and double-geometric payout implementations.
In such payout systems, the share's contribution to the payout (the weight for the score) follows the exponential function and decays by $de^{-dT}$ where $T$ is the time elapsed (recorded by the mining pool manager) since the share submission and $d$ is some constant for the decaying factor.
In other words, a share which has been submitted at time $x$ will have $e^{-d\Delta x}$ less payout value than the share submitted at time $x+\Delta x$ for any $x$ and any $\Delta x>0$. 
Because it decays by exponential, a share's average score is $\frac{1}{d}$. 
The payout is then distributed proportionally to the shares' contribution to the scores, i.e.,
the attacker's payout is the sum aggregate of the scores of the shares submitted by the attacker divided by the total sum aggregate of all the shares' scores.
The score for any miner $j$, denoted by $S_j$, is the sum-aggregate of the scores of the shares submitted by $j$.


For every block, there are multiple shares by design,
and the impact of SWH depends on the number of shares submitted in a round,
which number corresponds to $\| \vec{s} \|$ (where $\| \vec{x} \|$ is the number of elements of any vector $\vec{x}$).
We also define $\gamma$ to be the ratio between the block difficulty and the share difficulty, i.e., $\gamma = \frac{D_{\mbox{block}}}{D_{\mbox{share}}}$ where $D_{\mbox{block}}$ is the measure for difficulty for finding a block and is inversely proportional to the solution threshold $\tau_{\mbox{block}}$ (described in Section~\ref{sec:intro}) and $D_{\mbox{share}}$ is the share difficulty and inversely proportional to $\tau_{\mbox{share}}$.
$D_{\mbox{block}}$ is controlled by the blockchain system, 
e.g., Bitcoin adjusts $D_{\mbox{block}}$ according to the total computational power so that a block gets mined every ten minutes in expectation,
and $D_{\mbox{share}}$ is to increase the contribution-estimation samples to lower the variance, e.g., aggregating more samples converge better to the mean by law of large numbers.
$\gamma = \frac{D_{\mbox{block}}}{D_{\mbox{share}}}$ corresponds to the number of shares per blocks, i.e., for every block, a miner finds $\gamma$ shares on average.

SWH is analyzed within the pool since it unfairly increases the attacker's payout, the attacker's fraction of the victim pool reward.
We introduce the \emph{normalized payout} $\Gamma$,
which is the fraction of $f_j$ with respect to the aggregate-sum of the norm of all the elements of $f$ assuming that the miner $j$ is the attacker.
For example, if $\Gamma=1$, then the attacker earns all the pool reward and no other members within the pool receives the pool reward.
To simplify the analyses and focus on the intra-pool perspective, we also introduce an intermediate variable $\alpha'$ which is the attacker's computational power within the victim pool and $\alpha' = \alpha \tau$ (as defined in Section~\ref{subsec:computational_power_model}).
For example, the victim pool has a computational power of $\alpha'+\beta$ from both the attacker and the other miner members.

To quantify the attacker's reward, we define $c'$ as the attacker's withheld share's payout given that another miner within the victim pool broadcasts a block. 
In other words, given that the attacker withholds shares and that it detects the block submission from another miner within the victim pool, 
$c'$ quantifies the fraction of the payout which can be earned from the withheld shares if the attacker submits the shares at the time of the detection. 
$c'$ is analogous to $c$ from FAW in that it is random and dependent on the attacker's networking topology and environment, e.g., the attacker can increase $c'$ by optimizing routing and forwarding and regularly checking/maintaining the connectivity to the pool manager or, if the attacker is capable of launching a networking-based attack of eclipse attack on the pool manager~\cite{heilman2015}, $c \rightarrow 1$ and $c' \rightarrow 1$. 
However, in contrast to $c$, $c'$ also has a deterministic factor which is the stale share reward given by the mining pools, which can be motivated to do so since the ``stale'' shares solving the puzzle in the previous round can result in uncle rewards (stale block rewards). 
$c'$ is lower-bounded by such factor; if the mining pool provides $x$ reward for stale pools where $0 \leq x \leq 1$, then $c' \geq x$.
We investigate the impact of $c'$ on SWH and the attacker's reward in Section~\ref{subsec:sim_swh_reward}.

%

Appendix~\ref{appendix:payout_honest} analyzes the payout of the honest mining to provide an example case of using our mining-pool model. 
The following summarizes the additional variables used for the payout/reward analyses.
\newline \indent \indent $d$: $\; \; $ Decaying factor of the share's payout
\newline \indent \indent $S_j$: $\; \; $ Score for the miner $j$ 
\newline \indent \indent $\gamma$: \indent  Ratio of the block difficulty and share difficulty
\newline \indent \indent $\Gamma$: \indent  Payout 
to the attacker
\newline \indent \indent $\alpha'$: \indent  Attacker's computational power in the victim pool
\newline \indent \indent $c'$: \indent  Fraction/probability of payout earned from the withheld shares given that attacker detects another block within the victim pool 

\subsection{SWH Payout} 
\label{subsec:swh_payout}

Suppose the attacker knows when the block will get submitted ($t_B$).  
To maximize its reward, the share-withholding attacker submits all the shares right before $t_B$ (at $t_B-\epsilon$ where $\epsilon$ is small),
so that all shares experience no decay in their payout weight (score) and have the maximum payout weight of one.

While each share's score contribution individually is $d$ times greater than have the attacker behaved honestly and submitted the shares as they were found so that they experienced the decay, the final payout is distributed proportionally to the miner members' scores.
Therefore, share-withholding attacker $j$ increases its payout from $\Gamma_{\mbox{honest}} = \frac{\alpha'}{\beta+\alpha'}$ to $\Gamma_{\mbox{SWH}} = \mbox{E}\left[ \frac{S_j}{S_j + \sum_{i\neq j} S_i} \right]$ 
where $S_j$ is from the SWH shares which do not experience decay while $S_i$ are normal shares experiencing decay for the other miners $i$ where $i \neq j$.
The share history $\vec{s}$ and the payout function $f$ determine the attacker's payout, 
$\Gamma_{\mbox{SWH}}$.

\begin{theorem}
The expected payout of SWH is: 
\begin{eqnarray}
\Gamma_{\mbox{SWH}} \geq \sum_{y=1}^\infty \sum_{x=1}^y {y \choose x} \frac{x}{x+\frac{y-x}{d}} \cdot  \frac{\alpha'^{x} \beta^{y-x}}{(\alpha' + \beta)^y} \cdot \frac{(\gamma-1)^{y-1}}{\gamma^y}
\label{eqn:swh_reward_gain}
\end{eqnarray}
\label{thm:swh_payout}
\end{theorem}
\vspace{-0.2in}
\begin{proof}
The proof is in Appendix~\ref{appendix:swh_payout}. 
\end{proof}

\begin{corollary}
The expected payout of SWH is:
\begin{eqnarray}
\Gamma_{\mbox{SWH}} \approx  \frac{\alpha' d}{\alpha' d + \beta}
\label{eqn:reward_swh_approx}
\end{eqnarray}
\label{cor:swh_reward_approx}
\end{corollary}
\vspace{-0.2in}
\begin{proof}
The proof is in Appendix~\ref{appendix:swh_payout}. 
\end{proof}

Corollary~\ref{cor:swh_reward_approx} provides an approximation of the attacker's payout which is simpler than the iteration-based expression in Equation~\ref{eqn:swh_reward_gain} and is therefore easier to observe its behaviors with respect to its dependent variables.
Using Equation~\ref{eqn:reward_swh_approx}, the SWH attack impact in the payout grows with the attacker's power ($\alpha'$), decreases with the rest of the mining pool's power ($\beta$), and increases with the share score's decaying factor ($d$). 
We observe these behaviors in our simulations in Section~\ref{subsec:sim_swh_payout}.

\subsection{Share and Block Withholding: SWH-FAW and SWH-UBA} 
\label{subsec:swh_faw_reward}
FAW attack provides a concrete case when the attacker controls the timing of the block submissions (reactive to a third-party miner submitting a block, causing collision) and is therefore an opportune platform for share-withholding attack (SWH).
An FAW attacker submits the withheld block only when it discovers that there is a block getting propagated by a third-party pool.
In addition, a share-withholding attacker coupled with FAW attack (SWH-FAW) withholds the shares and submits them right before the withheld blocks.
%

Building on our analyses of FAW and UBA in Section~\ref{subsec:faw_recap}, 
the expected reward of SWH-FAW attack is the following, given the SWH attack payout $\Gamma_{\mbox{SWH}}$ (e.g., Equation~\ref{eqn:swh_reward_gain}):
{\footnotesize
\begin{eqnarray}
R_{\mbox{SWH-FAW}} = \frac{(1-\tau)\alpha}{1-\tau\alpha} + \Gamma_{\mbox{SWH}} \left( c' \frac{\beta}{1-\tau \alpha} + c \tau \alpha \frac{1- \alpha - \beta}{1-\tau \alpha} \right)
\end{eqnarray}}

Similarly, 
the reward of SWH-UBA attack is the following:
{\small
\begin{eqnarray*}
R_{\mbox{SWH-UBA}}
& \geq & \frac{(1-\tau) \alpha}{1-\tau\alpha}
+ \Gamma_{\mbox{SWH}} \left( \kappa \frac{(\tau \alpha)^2}{\beta + \tau \alpha} \cdot \frac{(1-\tau) \alpha}{1-\tau\alpha} 
+ c' \frac{\beta}{1-\tau \alpha}
+ \kappa \tau \alpha \frac{1-\alpha-\beta}{1-\tau \alpha} \right)
\label{eqn:reward_swh_uba_approx}
\end{eqnarray*}
}

\section{\rev{The Equilibrium Analysis}}

So far we assumed that there exist both rational miners and honest miners (protocol complying and no withholding), i.e., not all miners are rational,
because implementing rational miner requires the change in strategy and the update in the mining software from the default code, e.g., for Bitcoin or Ethereum.
In this section, we analyze the case where all the miners are rational as a miner's goal is to earn financial profit in general. We summarize our analysis in this section while Appendix~\ref{append:equilibrium} provides greater details. 
A rational miner equipped with FAW or UBA (the mining strategies described in Section~\ref{sec:related} except for BWH) yield the Nash equilibrium where the miners attack each other by launching FAW or UBA without Miner's Dilemma. 
If the rational miner can also join the mining pool with the greatest aggregate computational power (e.g., because it is an open pool), the rational miners congregate to the pool in Nash equilibrium, 
resulting in mining centralization~\cite{kwon2017,kwon2019}.
SWH only reinforces such equilibrium because it further amplifies the reward gain of the attacks beyond protocol compliance. 

\section{Simulation Analyses}
\label{sec:sim}

\rev{To analyze SWH, our model introduces environmental parameters ($\alpha,\; \beta$), attacker's control parameters ($\tau$), and the blockchain system control parameters ($\kappa, \lambda, \gamma, d$).
We focus our analyses from the attacker's perspective (observing the attacker's reward) and thus vary the attacker's parameters ($\alpha, \; \tau$) while using $\beta=0.24$, $\gamma = 2^5$, and $d=2^5$ (Appendix~\ref{appendix:sim_setup} describes how these system parameters are derived from modern cryptocurrency implementations). 
Appendix~\ref{subsec:sim_swh_payout} presents our simulation results for analyzing the SWH payout within a pool while Section~\ref{subsec:sim_swh_reward} studies the SWH reward performances and compare them with the existing schemes in FAW, UBA, and honest mining. 
}

\begin{figure*}[t]
\begin{center}
\subfigure[$c'$=0]{
	\includegraphics[width=0.31\textwidth]{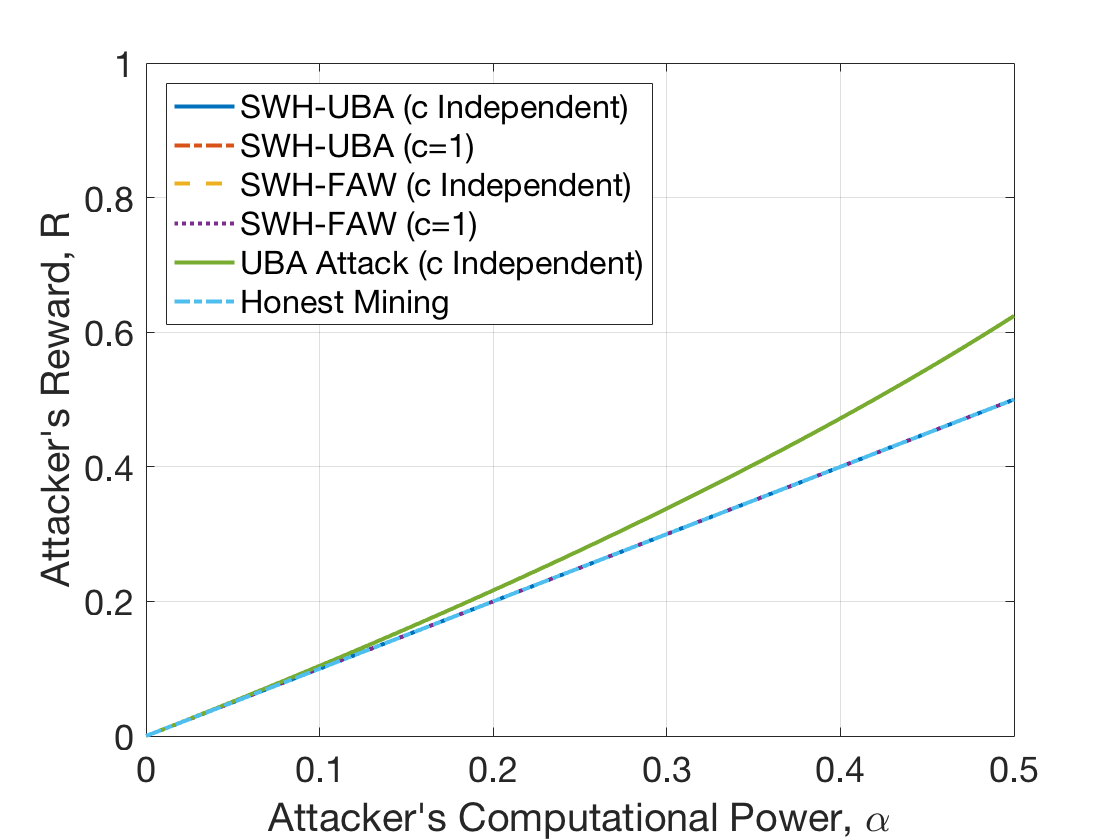}
       \label{fig:swh_reward_0}}
\hfill
\subfigure[$c'=\frac{1}{3}$]{
       \includegraphics[width= 0.31\textwidth]{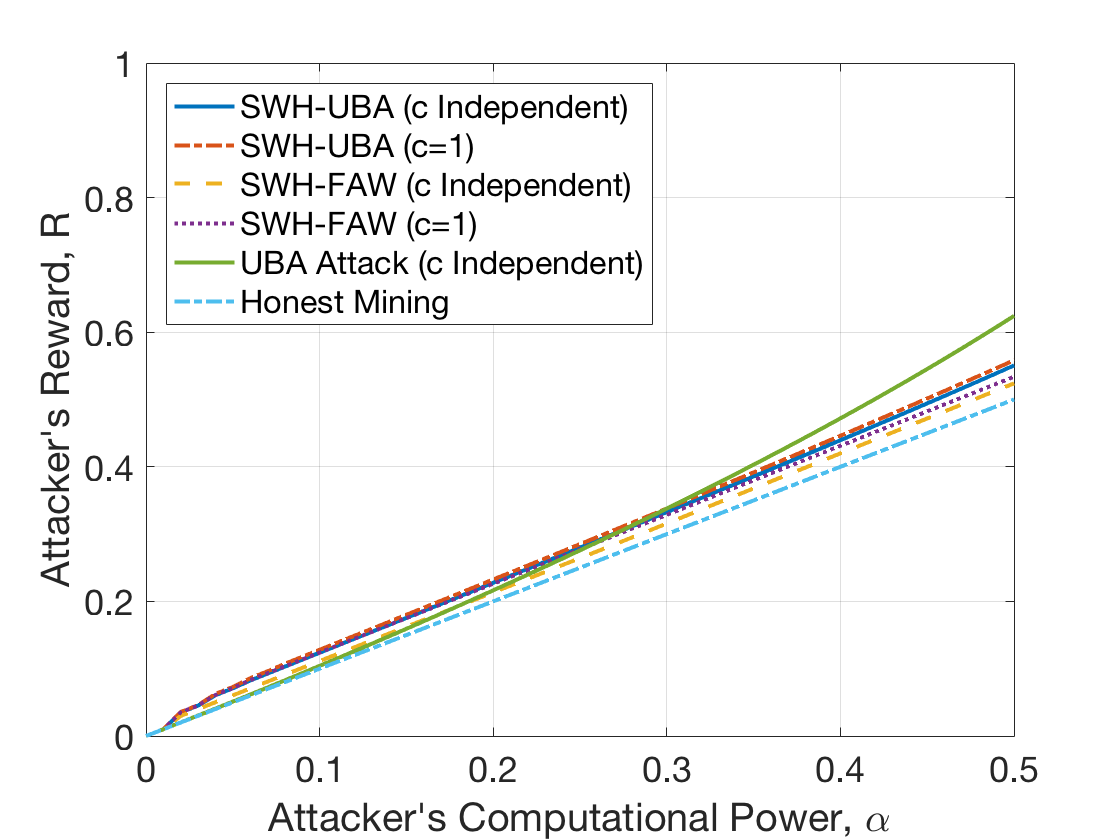}
       \label{fig:swh_reward_033}}
\hfill
\subfigure[$c'=1$]{
       \includegraphics[width= 0.31\textwidth]{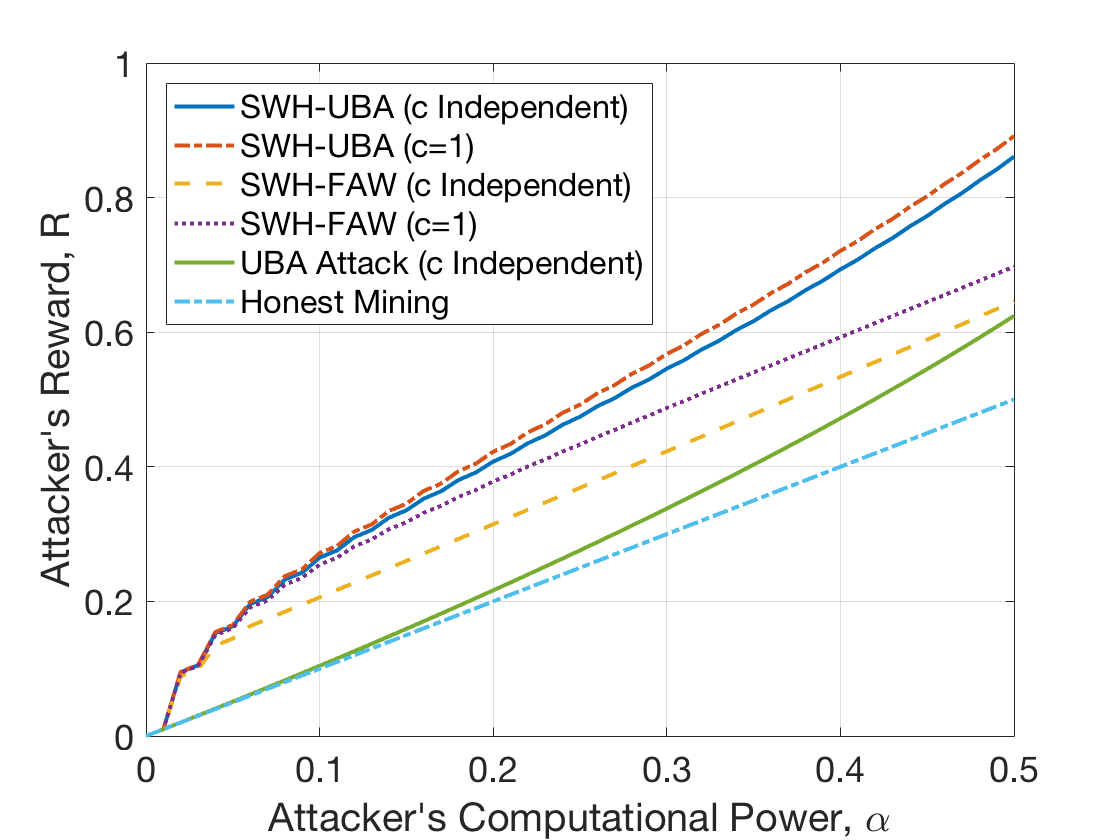}
       \label{fig:swh_reward_1}}
\caption{Reward comparison for SWH combined with block withholding threats (FAW and UBA), UBA only, and honest mining across different $c'$}
\label{fig:sim_swh}
\end{center}
\end{figure*}

\subsection{SWH Reward}
\label{subsec:sim_swh_reward}

Because SWH requires the block submission timing information as discussed in Section~\ref{sec:swh_scope},
we analyze when SWH attack is combined with UBA or FWA. 
%
Assuming that the attacker dynamically controls $\tau$,  
Figure~\ref{fig:sim_swh} compares the rewards of the attacks of combining SWH and UBA (SWH-UBA), combining SWH and FAW (SWH-FAW), UBA attack without SWH (UBA), and honest mining with no withholding (honest mining). 
For the non-SWH strategies, we focus on honest mining and UBA since UBA generalizes and outperforms FAW, which in turn outperforms the original BWH. 
For SWH-enabled attacks, Figure~\ref{fig:sim_swh} shows the worst-case for c (c=0 which yields the ``c independent'' reward) and the best-case for $c$ ($c=1$) in different curves;
any networking environment in between will yield performances between these two extreme reward performances. 

We study the attack impact/performances in different environments in $c'$ (which can be influenced by the mining pool implementations and the networking environments, as described in Section~\ref{subsec:model_decaying_payout}). 
Because it characterizes the general racing condition inside the pool (as opposed to between pools as $c$ does), $c'$ only affects SWH and does not affect the block-withholding based attacks of UBA and FAW. 
The reward performances behave differently according to the magnitude of $c'$. 
When $c'$ is low, the SWH-enabled attacks of SWH-UBA and SWH-FAW force the optimal attacker strategy to become that of either UBA (without SWH) or honest mining (no withholding of blocks and shares), as seen in Figure~\ref{fig:swh_reward_0}. 
As is seen in Figure~\ref{fig:gain_vs_cprime} showing the reward gain,
the SWH-UBA attack becomes better than honest mining when $c'\geq0.153$ and better than UBA-only attack when $c'\geq0.201$ when $\alpha=0.1$; 
as the attacker's computational capability $\alpha$ increases, the SWH-UBA attack outperforms honest mining and UBA attack with even lower $c'$. 
However, such low $c'$ is unlikely for an attacker launching SWH as described in Section~\ref{subsec:model_decaying_payout}. 

For other environments with $c'$ of intermediate magnitudes (e.g., Figure~\ref{fig:swh_reward_033}), 
to maximize its reward, the attacker will choose between SWH-UBA and UBA depending on its power capability $\alpha$.
A lower-power attacker (smaller $\alpha$) will choose SWH-UBA while a power-capable attacker (large $\alpha$) will choose UBA attack (without SWH) for maximizing its reward, since the SWH-UBA attack increases more rapidly with $\alpha$ than UBA. 
According to Figure~\ref{fig:gain_vs_cprime}, the SWH-UBA attacker will opt to UBA-only when $c' \leq 0.201$ at $\alpha=0.1$, but this changeover point for the optimal attack strategy will occur at greater $c'$ with greater $\alpha$ capabilities (e.g., UBA-only attack is optimal when $c' \leq 0.304$ if $\alpha=0.24$). 
In other words, with smaller power capabilities ($\alpha$), the attacker will more likely choose SWH-UBA over UBA without SWH to optimize its reward capacity with varying pool/networking environments ($c'$). 
Lastly, for large $c'$ (e.g., Figure~\ref{fig:swh_reward_1}), SWH-UBA attack outperforms UBA attack regardless of $\alpha$. 







\section{Conclusion}
\label{sec:conclusion}
Blockchain uses PoW to achieve consensus in cryptocurrencies and other permissionless applications. 
From real-world blockchain and mining system implementations, we identify and analyze the blockchain system components causing vulnerabilities for unfair reward exploitation 
and introduce the SWH threat. 
SWH attacks the victim mining pool and increases the payout within the pool, 
in contrast to the block-withholding based threats which consider the inter-pool dynamics and increase the attacker's main pool reward by sabotaging the victim pool. 
If launched along with the block-withholding threats, SWH is effective 
in gaining unfair reward advantage at the expense of the other protocol-complying honest miners and is aligned with the incentives of rational and financially driven miners in general. 
Since the attack requirements for launching such threat is comparable to that of BWH, FAW, or UBA 
(and there are already reported incidents of BWH in real world), 
we expect the more impactful SWH-UBA or SWH-FAW to occur against vulnerable systems in practice, such as Slush Pool and other mining pools implementing decaying payout. 
We intend to inform the blockchain R\&D community of the realistic and impactful SWH threat 
and discuss about the potential countermeasure directions in Appendix~\ref{subsec:potential_countermeasures} to facilitate further research and development. 


\section*{Acknowledgement}
This research is supported in part by Colorado State Bill 18-086.
We would also like to thank the anonymous reviewers for their helpful feedback. 

\bibliographystyle{IEEEtranS_ndss}   
\bibliography{blockchain_2}  


\begin{subappendices}
\renewcommand{\thesection}{\Alph{section}}%

\section{\rev{UBA Reward Analysis}}
\label{appendix:uba_reward}

\rev{UBA builds on FAW but adapts its strategy so that the attacker submits the withheld block in Event $B$ and earns additional rewards in Event $C$. 
The expected reward for UBA ($R_{\mbox{UBA}}$) is: 
{\footnotesize 
\begin{eqnarray}
R_{\mbox{UBA}} 
 \geq & \frac{(1-\tau) \alpha}{1-\tau\alpha} &
+  \frac{\tau \alpha}{\beta + \tau \alpha}  \Bigg( \kappa \frac{(\tau \alpha)^2}{\beta + \tau \alpha} \cdot \frac{(1-\tau) \alpha}{1-\tau\alpha} \nonumber \\
&  & + (1 + \kappa \tau \alpha) \frac{\beta}{1-\tau \alpha} 
+ \kappa \tau \alpha \frac{1-\alpha-\beta}{1-\tau \alpha}   \Bigg) 
\label{eqn:reward_uba_approx}
\end{eqnarray}
}
where $\kappa<1$ is the partial reward to the uncle blocks. 
The right-hand side assumes that there is only one uncle block reward, which yields the inequality. If there are more uncle block rewards, e.g., Ethereum's GHOST algorithm, then the reward increases beyond the right-hand side in Equation~\ref{eqn:reward_uba_approx}.  }

\section{Payout Analysis for Honest Mining}
\label{appendix:payout_honest}

To provide an example case for using the model in Section~\ref{subsec:model_decaying_payout}, we analyze the case of honest mining.
We assume decaying payout and each share's score for payout is $\frac{1}{d} = \frac{1}{\gamma}$ in expectation since $\gamma$ is the expected number of shares per block if the mining pool wins the block (if the mining pool collectively does not find a block, then the payout becomes zero regardless of the share scores). 
Therefore, $d = \gamma$; while we focus on $d = \gamma$, we analyze the impact of varying $d$ in Section~\ref{subsec:sim_swh_payout}.

An attacker behaving honestly receives a reward of $ (\alpha' + \beta) \cdot \frac{1}{\gamma} \cdot \frac{\alpha'}{\alpha' + \beta} \gamma = \alpha'$ in expectation
where $\alpha' + \beta$ corresponds to the probability that the pool wins the block (against the other pools), and $\frac{1}{\gamma}$ and $\frac{\alpha'}{\alpha' + \beta} \gamma$ correspond to the reward per share and the number of shares (the total number of shares is $\gamma$ in expectation), respectively, given that the pool won the block. 
The resulting expected reward for the attacker behaving honestly is $\alpha'$, which is its computational power invested on the victim pool and agrees with 
Equation~\ref{eqn:reward_honest}.

With no SWH, the attacker's payout is 
$\Gamma_{\mbox{honest}} = \frac{\alpha'}{\alpha'+\beta} = \frac{\tau\alpha}{\tau\alpha+\beta}$ in expectation,
agreeing with our analyses in Section~\ref{sec:mining_game}. 



\section{SWH Payout Analyses: Proofs of Theorem and Corollary}
\label{appendix:swh_payout}

In this appendix section, we prove Theorem~\ref{thm:swh_payout} and Corollary~\ref{cor:swh_reward_approx} in Section~\ref{subsec:swh_payout}. 

\setcounter{theorem}{2}
\setcounter{corollary}{0}

\vspace{0.08in}
\begin{theorem}
The expected payout of SWH is: 
\begin{eqnarray*}
\Gamma_{\mbox{SWH}} \geq \sum_{y=1}^\infty \sum_{x=1}^y {y \choose x} \frac{x}{x+\frac{y-x}{d}} \cdot  \frac{\alpha'^{x} \beta^{y-x}}{(\alpha' + \beta)^y} \cdot \frac{(\gamma-1)^{y-1}}{\gamma^y}
\end{eqnarray*}
\end{theorem}
\vspace{0.08in}

\begin{proof}
The expected payout of the miner $j$ launching SWH with a power of $\alpha'$ is: 
\begin{eqnarray}
\Gamma_{\mbox{SWH}} 
& = & \mbox{E}\left[\frac{S_j}{S_j + \sum_{i\neq j} S_i} \right] \nonumber \\
& = & \sum_{y=1}^\infty \sum_{x=1}^y \mbox{E} \left[ \left. \frac{x}{x+\sum_{i\neq j} S_i} \right\vert S_i = x, \|\vec{s}\| = y \right] \cdot \mbox{Pr}[S_j = x \; \bigg| \|\vec{s}\| = y] \cdot \mbox{Pr}[\|\vec{s}\|=y]   \nonumber \\
& = & \sum_{y=1}^\infty \sum_{x=1}^y \mbox{E} \left[ \left. \frac{x}{x+\sum_{i\neq j} S_i} \right\vert S_i = x, \|\vec{s}\| = y \right]  \nonumber \\
& & \cdot {y \choose x} \left(\frac{\alpha'}{\alpha'+\beta} \right)^x \left(\frac{\beta}{\alpha'+\beta} \right)^{y-x} \cdot \left( 1-\frac{1}{\gamma} \right)^{y-1} \frac{1}{\gamma}   \nonumber \\
& \geq & \sum_{y=1}^\infty \sum_{x=1}^y \frac{x}{x+\mbox{E}\left[ \sum_{i\neq j} S_i \right]} \cdot {y \choose x} \left(\frac{\alpha'}{\alpha'+\beta} \right)^x \left(\frac{\beta}{\alpha'+\beta} \right)^{y-x} \cdot \left( 1-\frac{1}{\gamma} \right)^{y-1} \frac{1}{\gamma}   \nonumber \\
& = & \sum_{y=1}^\infty \sum_{x=1}^y {y \choose x} \frac{x}{x+\frac{y-x}{d}} \cdot  \frac{\alpha'^x \beta^{y-x}}{(\alpha' + \beta)^y} \cdot \frac{(\gamma-1)^{y-1}}{\gamma^y}  \nonumber 
\end{eqnarray}
${y \choose x}$ is the binomial coefficient, i.e., ${y \choose x} = \frac{y!}{x!(y-x)!}$, 
and E[$X|Y$] and Pr[$X|Y$], respectively, are the expected value and the probability of event $X$ given event $Y$ for some events $X$ and $Y$.
The second equality is from the Bayes' rule (conditioned on the attacker's final score $S_j$ and the number of shares in the round $\vec{s}$),
and the third equality is derived because $S_j$ has a binomial distribution with $y$ trials and $\frac{\alpha'}{\alpha'+\beta}$ probability (in each share, there is a $\frac{\alpha'}{\alpha'+\beta}$ chance that the share is being submitted by an attacker), given $\|\vec{s}\| = y$,
and $\vec{s}$ has a geometric distribution with the parameter/probability of $\frac{1}{\gamma} = \frac{D_{\mbox{share}}}{D_{\mbox{block}}}$ (the last share in $\vec{s}$ is the one that is also a block).
The inequality is due to Jensen's Inequality and that $\mbox{E}\left[\frac{S_j}{S_j + \sum_{i\neq j} S_i} \right]$ is convex with $S_j + \sum_{i\neq j} S_i$ given $S_j$.
Finally, linear algebra yields the final equality, 
which provides a lower bound on the attacker $j$'s payout.
\end{proof}
\vspace{0.08in}

\vspace{0.08in}
\begin{corollary}
The expected payout of SWH is:
\begin{eqnarray*}
\Gamma_{\mbox{SWH}} \approx  \frac{\alpha' d}{\alpha' d + \beta}
\end{eqnarray*}
\end{corollary}
\vspace{0.08in}

\begin{proof}
We approximate the gain of SWH with respect to honest mining by taking 
$\mbox{E}\left[\frac{S_j}{S_j + \sum_{i\neq j} S_i} \right] \approx \frac{\mbox{E}[S_j]}{\mbox{E}[S_j] + \mbox{E}[ \sum_{i\neq j} S_i]} $.
\begin{eqnarray}
\Gamma_{\mbox{SWH}} & \approx & \frac{\mbox{E}[S_j]}{\mbox{E}[S_j] + \mbox{E}[ \sum_{i\neq j} S_i]} \nonumber \\
& = & \frac{\frac{\alpha'}{\alpha'+\beta}\gamma}{\frac{\alpha'}{\alpha'+\beta}\gamma + \frac{\beta}{\alpha'+\beta} \gamma \frac{1}{d}} \nonumber \\
& = &  \frac{\alpha' d}{\alpha' d + \beta} \nonumber
\end{eqnarray}
$\mbox{E}[S_j] = \frac{\alpha'}{\alpha'+\beta}\gamma$ because the expected number of share for the attacker $j$ is $\frac{\alpha'}{\alpha'+\beta}\gamma$ and each share is weighted by one because of no decay,
and $\mbox{E}[ \sum_{i\neq j} S_i] = \frac{\beta}{\alpha'+\beta} \gamma \frac{1}{d}$ because the expected number of share fore the rest of the miners excluding $j$ is $\frac{\beta}{\alpha'+\beta} \gamma$ and each share's score contribution is decayed by $\frac{1}{d}$ on average.
\end{proof}
\vspace{0.08in}


\section{The Equilibrium Analysis}
\label{append:equilibrium}

In this section, we analyze the case where all the miners are rational as the miner's goal is to earn financial profit. 
We build on prior research in the mining strategies described in Section~\ref{sec:related} for our analysis and corroborate with the prior research yielding that rational miners congregate to the mining pool with the greatest computational power resulting in mining centralization~\cite{kwon2017,kwon2019}, i.e., SWH only reinforces such equilibrium analyses.  
(While such analysis is valid, Appendix~\ref{subsec:potential_countermeasures} discusses a potential measure introducing a distributed mining pool, as opposed to having a centralized pool manager, to mitigate the centralization issue.)

\rev{FAW, building on selfish mining and BWH, forgoes the miner's dilemma~\cite{eyal2015}, i.e., the Nash equilibrium of the FAW-capable miners do not result in the suboptimal tragedy-of-the-commons (where collaboration and coordination, as opposed to each miner's adopting their unilateral strategies, would have provided better performances for all the miners involved). As a result, if there is no restriction in joining any pool, the rational miners join and congregate to the mining pool with greater computational power and attack the other pools until no rational miner is left in the other pools. 
UBA further takes advantage of the rewards for the uncle blocks to further reinforce such Nash equilibrium. 
(In contrast, for BWH, because of the miner's dilemma, the Nash equilibrium is to leave the mining pools and mine directly in the blockchain's consensus protocol.)
SWH, requiring FAW or UBA for incentive compatibility (greater reward than honest mining), only increases the reward beyond FAW and UBA and utilizes an orthogonal channel for the reward gain because, while FAW and UBA increases the chance of the attacker's main pool of winning the reward, SWH increases the attacker's reward payout within the mining pool. 
Section~\ref{sec:swh_scope} establishes the blockchain- and pool-settings for SWH to be relevant for rational strategy for reward gain, and Section~\ref{subsec:sim_swh_reward} analyzes the attacker's power and networking conditions which make SWH more profitable than launching UBA or FAW only. 
Because SWH provides further reward gains for UBA/FAW, 
it reinforces the Nash equilibrium of UBA/FAW, i.e.,
the rational miners equipped with the SWH strategies are even more incentivized to congregate to the pool with the greatest mining power. 
On the other hand, if there is a mechanism to control the joining of the pool (e.g., some closed pools require registration but these pools are rarer because the miner gives up the cryptocurrencies' permissionless and anonymization properties) and the greater-power pool uses such mechanism, 
then the pools launch SWH and UBA/FAW against each other and the larger pool wins by earning greater reward gains than the smaller pools. 
}



\section{Simulation Setup and Parameters}
\label{appendix:sim_setup}

This section explains the simulations setup and the parameter choices to characterize the blockchain system and the victim pool system under attack.

Our blockchain system simulation setup is influenced by modern blockchain implementations.
%
For the pool system, $\beta=0.24$, which value corresponds to the strongest mining pool in real-world mining at the time of this manuscript writing~\cite{blockchain_info_hashrate}.
The attacker attacking the stronger pool as its victim pool (as opposed to a weaker pool of $\beta \rightarrow 0 $) provides greater reward and is aligned with its incentive, which we verify in our simulations and agree with previous literature~\cite{kwon2017,bag2016}.
The pool difficulty of the victim pool corresponds to $\gamma = 2^5$, i.e., the share difficulty is 5 bits (32 times) less than the block difficulty.
This falls within the typical range of $\gamma$~\cite{gamma2012}, which parameter provides a pool-system-controllable tradeoff between networking/bandwidth and reward variance.
The decaying factor for the pool's reward payout is $d=2^5$; we have $d = \gamma = 2^5$, so that the shares scores decay by $e^{-dT}$ in time $T$ and the scores add up to one in expectation, as described in Section~\ref{subsec:model_decaying_payout}.
These parameters are fixed unless otherwise noted
(we vary the variables to analyze the dependency and the impacts).

We also consider the 51\% attack where the attacker can fully control the blockchain if the attacker's computational power exceeds the 50\% of the network's;
in our context, the attacker can conduct withholding-based selfish mining to reverse the transactions/blocks on the chain
and to waste the other miner's computational resources on blocks that the attacker can reverse and make stale.
Therefore, we limit our analyses to $0 \leq \alpha \leq 0.5$ (the attacker is capable of 51\% attack if $\alpha>0.5$) in addition to the constraint of $\alpha + \beta \leq 1$ from the definitions of $
\alpha$ and $\beta$.

%
\begin{figure}[t]
\begin{center}
\subfigure[Payout $\Gamma$ with respect to $\gamma$]{
       \includegraphics[width= 0.31\columnwidth]{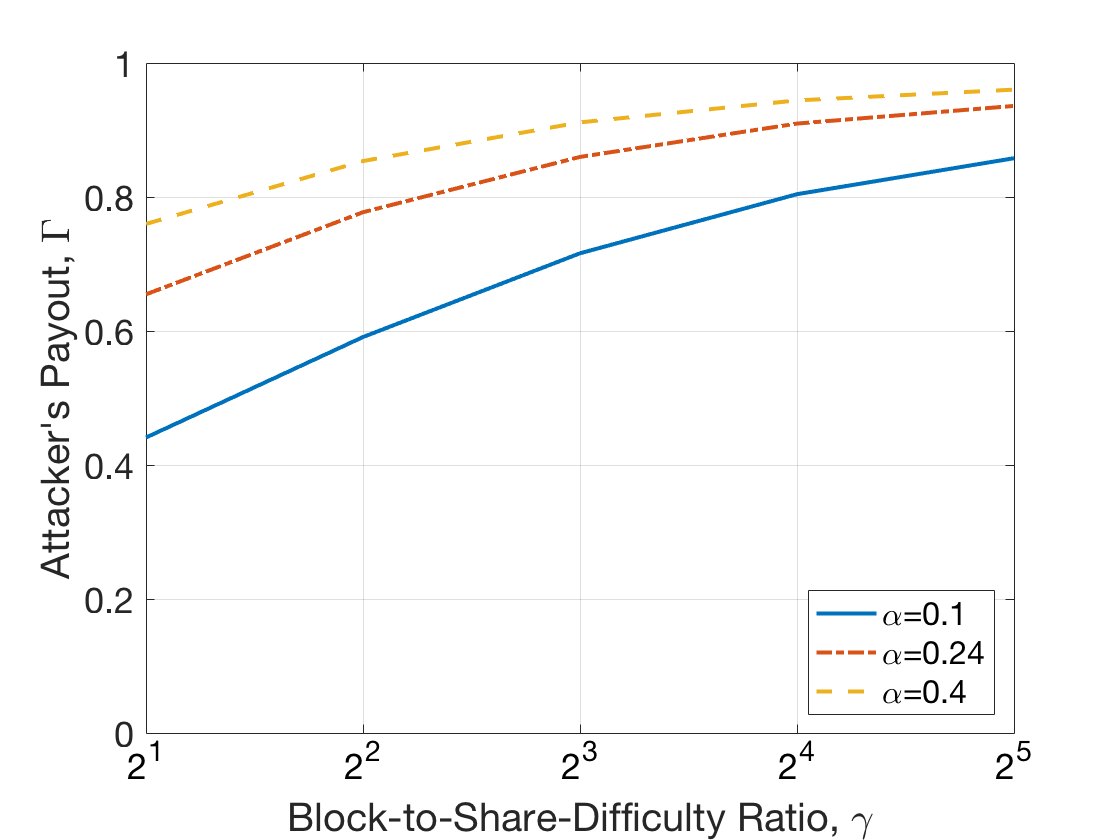}
       \label{fig:swh_Gamma_gamma}}
\subfigure[Payout $\Gamma$ with respect to the share's decaying factor $d$]{
       \includegraphics[width= 0.31\columnwidth]{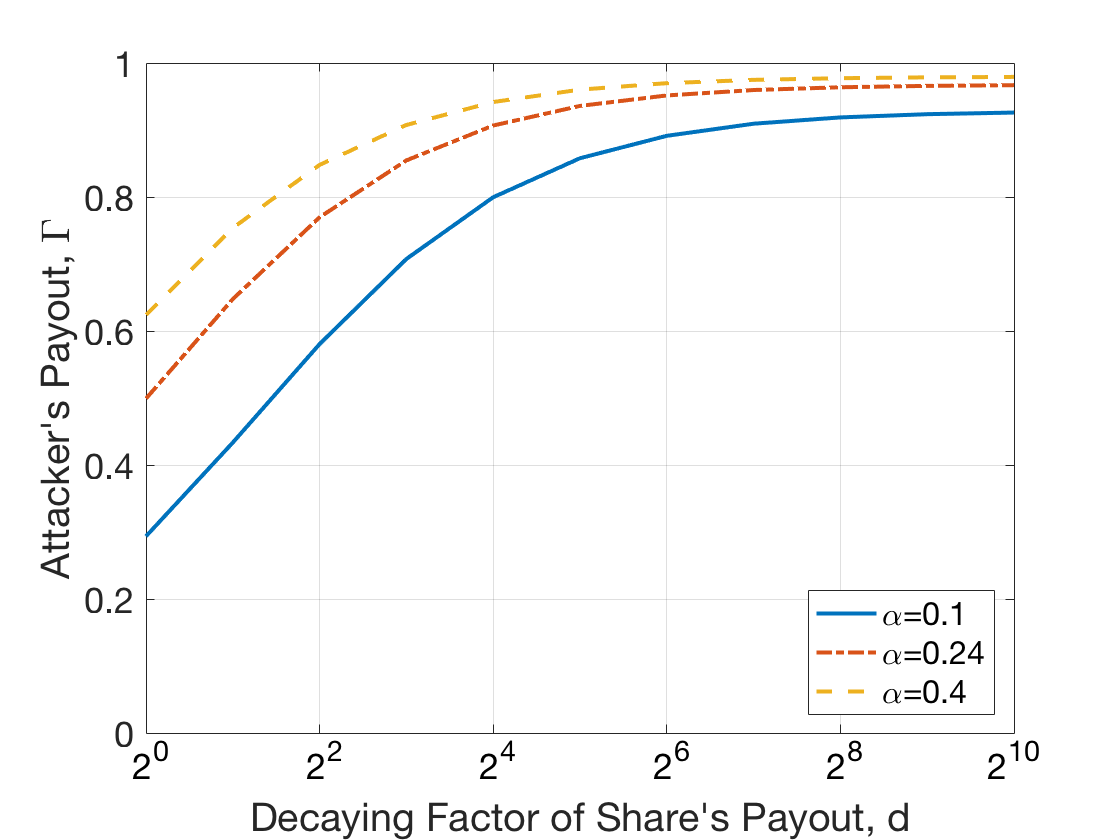}
       \label{fig:swh_Gamma_d}}
\subfigure[SWH-UBA's reward gain analyses assuming worst-case networking of $c=0$]{
       \includegraphics[width= 0.31\columnwidth]{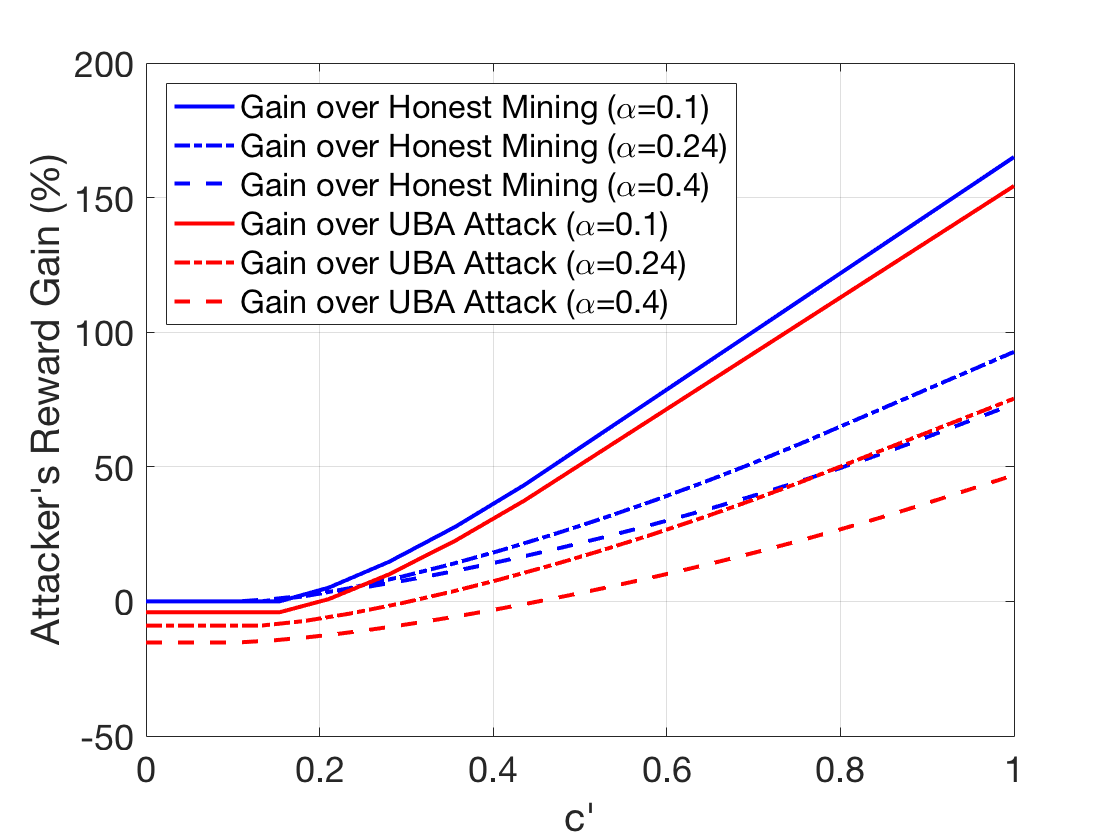}
       \label{fig:gain_vs_cprime}}
\caption{SWH payout analyses (Figures~\ref{fig:swh_Gamma_gamma} and~\ref{fig:swh_Gamma_d}) and SWH-UBA's reward gain analyses (Figure~\ref{fig:gain_vs_cprime})}
\label{fig:sim_swh_payout}
\end{center}
\end{figure}



\section{SWH Payout}
\label{subsec:sim_swh_payout}

SWH yields an unfair payout advantage to the attacker.
Assuming that the victim pool uses a decaying-payout scheme for distributing the pool reward
and that the attacker can submit the shares on time (i.e., $c'=1$), we study the payout advantage of the SWH attacker.
Corollary~\ref{cor:swh_reward_approx} in Section~\ref{subsec:swh_payout} provides an approximation for the attacker's payout in Equation~\ref{eqn:reward_swh_approx}, which can characterize the payout's dependency on $\alpha$, $\beta$, and $d$.
For example, the attacker's payout increases with $d$, the decaying factor of the share's payout, as is also shown in simulations in Figure~\ref{fig:swh_Gamma_d}.
We observe such behaviors in our simulations.
We also summarize our findings here when comparing the approximated payout in Equation~\ref{eqn:reward_swh_approx} and the more precise payout in Equation~\ref{eqn:swh_reward_gain}.
The approximation generally provides a higher payout than the more precise lower bound, 
and the difference ranges from 0\% (when the attacker's power on the victim pool is very low) to 21.1\%.
Since the difference is significant between the approximation and the lower bound we identified, 
we use the lower bound in Equation~\ref{eqn:swh_reward_gain} to quantify the attacker's payout rather than the approximation in Equation~\ref{eqn:reward_swh_approx}.
In addition, 
the resulting payout from launching SWH significantly outperforms honest mining (where the attacker does not withhold shares and submits them as soon as they are found);
while there is no payout difference between SWH and honest mining when $\alpha=0$, the difference quickly becomes the maximum of 0.5813 at $\alpha=0.06$ and then monotonically decreases and becomes 0.292 at $\alpha=0.5$.

The block-to-share difficulty ratio $\gamma$ establishes how often the shares occur for every block/round;
in expectation, there are $\gamma$ shares per block.
The greater the number of shares per round the greater the impact of the share-withholding attack, since the share-withholding attack can occur for every shares, as is seen in Figure~\ref{fig:swh_Gamma_gamma}.
In practice, mining pools typically control $\gamma$ and $d$ together so that they are aligned/correlated with each other, i.e., as there are more shares ($\gamma$ increases), the payout per share decreases more quickly ($d$ increases).
However, to isolate the effect of $d$ from $\gamma$,
Figure~\ref{fig:swh_Gamma_d} plots the payout $\Gamma$ with respect to $d$ while fixing $\gamma=2^5$
and shows that the attacker's payout $\Gamma$ increases with $d$.

\section{\rev{Discussions About Potential Countermeasures}}
\label{subsec:potential_countermeasures}

\rev{While this paper focuses on the discovery and the analyses of SWH attack, we discuss potential countermeasures for future work in this section. }

\subsubsection{\rev{Behavior-Based Detection of Withholding Threats}}
\label{subsec:countermeasure_detection}
\rev{The withholding-based threats, including SWH, result in abnormal reward behaviors. 
For example, SWH decreases the variance and the entropy of the share arrivals.
Such phenomenon can be sensed and measured for attack detection, which can be then used for mitigation purposes. }
\rev{While we identify behavior-based detection as promising, we do not recommend relying on identity-based detection and mitigation such as blacklisting public keys or IP addresses because, in permissionless environment, there is no identity control/registration and it is cheap for the attacker to generate multiple identities (Sybil attack).} 


\subsubsection{\rev{Payout and Reward Function Control}}
\rev{Controlling the payout and reward functions, 
for example, the system parameters $\kappa, \; \lambda, \; \gamma, \; d$ in our model, provides a low-overhead countermeasure because it requires the changes in the pool manager only and is backward-compatible to the rest of the miners' software. 
Prior research~\cite{bag2016,AWRS2019} distinguish between block submissions and share submissions (different weights) against BWH attack. Such approach can not only be used for BWH but also for FAW and UBA, which in turn defends against SWH because SWH relies on FAW or UBA for its incentive compatibility (SWH requires the attacker to have some/probabilistic information of the block arrivals). }

\subsubsection{\rev{Oblivious Share}}
\rev{Similar in purpose to the oblivious transfer protocols and building on commit-and-reveal approach, 
\emph{oblivious share} deprives the miner of the knowledge of whether it is a block or a share until it submits them~\cite{rosenfeld2011,2ppow_2014}.
The attacker therefore cannot dynamically adopt the withholding-based threats which require distinguishing the share and the block before submission. 
While effective against the wittholding-based attacks, such approach requires a protocol change (including an additional exchange between the mining pool manager and the miners) and is not backward compatible (does not work with the existing system unless the protocol change/update is made)~\cite{kwon2017,luu2015}, causing protocol/communication overheads and making such schemes undesirable for implementation to the blockchain network (which includes closed pools and solo miners, free of withholding vulnerabilities and thus lacking the incentives for such addition and change). }

\subsubsection{\rev{Unified Distributed Mining Pool}}
\label{subsec:one_pool_sol}
\rev{
To have 
all miners join one distributed mining pool eliminates the notion of sabotaging/victimizing another pool. 
A useful platform for this can be distributed mining pools, e.g., SmartPool~\cite{smartpool2017} and P2Pool, which eliminates the centralized mining pool manager and replaces it with a distributed program/computing,
motivated to make the blockchain computing more decentralized without the reliance on trusted third party (the mining pool manager in this case)~\cite{gervais2014,bonneau2015}.
Since the mining pool is distributed, the mining and the consensus protocol does not have the centralization issue, e.g., there is no centralized pool manager capable of controlling the rewards or blocking/nullifying a share or a block. 
In fact, the authors of SmartPool~\cite{smartpool2017} envisions that their platform can be used to unify the mining pools citing that the elimination of the mining pool fee (charged by the centralized mining pool manager for their services) and the reduced variance (compared to independent mining) will attract incentive-driven miners.
However, despite such desirable properties (and even if the claimed superior performance is true), 
such approach is a radical solution and it can be difficult to enforce the change in behaviors in the miners and have all miners mine at the designated pool, especially for an existing blockchain implementation with the existing miners having already joined a pool. 
Enforcing such pool restriction for the miners is not backward-compatible to the existing miners and can also be controversial since it can be viewed as a violation of the freedom of the miners. }


\end{subappendices}


\end{document}